\documentclass{cernphprep} 
                           
\usepackage{graphicx} 
\usepackage{verbatim} 
 
 \newcommand{\npt}{$p_T$}

 \newcommand{\Gc}{GeV/$c$}
 \newcommand{\Jpsi} {J/$\psi$}

 \newcommand{\nY}{$\Upsilon$}

\begin{document}

\begin{titlepage}
\PHdate{21 May 2015}

\title{Toward the Limits of Matter: Ultra-relativistic nuclear collisions at CERN}

 \begin{abstract}
Strongly interacting matter as described by the thermodynamics of QCD undergoes a phase transition, from a low temperature hadronic medium to a high temperature quark-gluon plasma state. In the early universe this transition occurred during the early microsecond era. It can be investigated in the laboratory, in collisions of nuclei at relativistic energy, which create "fireballs" of sufficient energy density to cross the QCD Phase boundary. We describe 3 decades of work at CERN, devoted to the study of the QCD plasma and the phase transition. From modest beginnings at the SPS, ultra-relativistic heavy ion physics has evolved today into a central pillar of contemporary nuclear physics and forms a significant part of the LHC program.
 \end{abstract}

\vfill

\begin{Authlist}
J\"urgen Schukraft
\Instfoot{a1}{PH Division, CERN, CH-1211 Geneva 23, Switzerland}

Reinhard Stock
\Instfoot{a2}{Institute of Nuclear Physics, and FIAS, Max von Laue Strasse 1, D-60438 Frankfurt/Main, Germany}

\end{Authlist}
\vfill
\Submitted{To appear in ``60 Years of CERN Experiments and Discoveries", Editors H. Schopper and L. di Lella, World Scientific Publishing, Singapore, 2015}

\end{titlepage}

 \section{Strongly interacting matter}
 We recall here the development of a novel research field at CERN, devoted to the phases and phase structure of matter governed by the strong fundamental force. Its proper field theory was discovered around 1970: Quantum Chromodynamics (QCD) addresses the fundamental interactions of elementary quarks, as mediated by gluons. Importantly, the gluons carry strong charges themselves, unlike the uncharged photons that mediate the QED interaction. Thus QCD is a much more complicated theory, mathematically. Both these field theories constitute a part of the modern Standard Model of elementary interactions. They thus enter electrons, photons, quarks and gluons into our inventory of elementary particles. Their predicted properties have been meticulously studied and confirmed by decades of particle physics research.
 
 Now, our interest here is not the study of elementary QCD collisions, but of extended matter governed by the strong interaction, from a partonic plasma state to protons and nuclei, and to neutron stars. Matter has arisen at an extremely early stage of the cosmological evolution, and is thus a part of the Cosmological Standard Model (developed in parallel to the Standard Model of elementary interactions). The universe has gone through successive, distinct stages of matter composition. From attoseconds to early microseconds the expanding cosmos was governed by a plasma 'fireball' era, and composed of elementary quarks, gluons, electrons etc. The Big Bang matter was conductive toward, both, electric and colour charge currents, but structureless otherwise. Our research field of QCD matter begins with this era.
 
 Cooling provokes structure formation via phase transitions. Note that they occur in macroscopic volumes only: we talk about phase transitions of extended matter. More precisely: although the transformations occur due to inter-quark neutralization and binding effects taking place at the microscopic scale, it is their collective synchronisation by macroscopic thermodynamic conditions such as density and temperature that leads to the emergence of states and phases of matter. The laws of thermodynamics, plus the microscopic intrinsic features of the carriers of degrees of freedom, and their interaction, constitute a characteristic phase diagram, with boundary lines between phases, in a plane of temperature and density.
 
 What is the phase diagram of a macroscopic volume of QCD matter consisting of quarks and gluons?
 
 This question represents the goal of the research field, addressed here. In other words we thus ask for the thermodynamics of QCD. Thus, in marked contrast to particle physics focus on the elementary properties of partons, we wish to know what will go on in, say, a cubic meter of quark-gluon plasma, of primordial cosmological vintage, once it cools down to nucleons, or even recompresses in the interior of neutron stars. This is a deeply non-trivial question, concerning partons and hadrons with non-perturbative, in-medium modified interactions. 
 
 In the Big Bang cosmological evolution expansive cooling descends, from an initially arbitrarily high energy density of partons without structure because QCD bound states (hadrons) could not form, down to a much lower temperature where, in turn, no free partons could exist, but hadrons. The partons exhibit a 'horror vacui' according to QCD that results from the fact that the force-mediating gluons carry strong colour charges themselves; emitted from a quark they have to be absorbed by another quark. QCD tells us that the quarks have to stay 'confined', at low energy density, within colour-neutral bound states, the hadrons. Indeed the cosmic evolution has left us with $10^{78}$ protons and neutrons but free quarks have never been observed. Thus there must have occurred a phase transition from partonic to hadronic matter, at some intermediary density. We cannot read the corresponding critical temperature directly off the present cosmological matter.
 
 Two considerations might give insight before starting with experiments. As hadrons must form in a phase transition from quark matter plasma to bound quark clusters, i.e. in some form of condensation, we should expect that the protons interior energy density should closely resemble that of the medium they condense from. Now we know that the proton interior energy density amounts to about 1 GeV per cubic Fermi. Second, the CERN physicist R. Hagedorn had demonstrated long before the advent of QCD that the hadronic world must have a temperature limit, of about 165 MeV. Remarkably, in a gas of partons one could estimate that the energy density at this temperature roughly corresponds to the above 1 GeV/fm$^3$!
 Now one infers from the Einstein-Friedman equations, long known to describe the space-time development that this density corresponds to the cosmological period in the early microsecond time domain. One cubic meter of matter at this time had ten times more mass than the entire Nanga Parbat mountain range.
 
 \section{QCD matter research: gaining confidence}
 The perspective of a phase transition in QCD matter, sketched above, reflects the state of theoretical physics at around 1976 when Stephen Weinberg~\cite{sweinberg} wrote his famous book 'The First Three Minutes'. Its last chapter addresses the early stages of the cosmological expansion, and it is his point that the universe was a thermal fireball evolving in thermodynamic equilibrium. Hot, thermal QCD had just been developed by Shuryak~\cite{Shuryak1}, Baym~\cite{Baym:1976yu} and Kapusta~\cite{Kapusta:1979fh}, introducing the name 'Quark-Gluon Plasma'(QGP) for a free parton 'gas', following a first suggestion by Collins and Perry~\cite{Collins:1974ky}. They had referred to the just-discovered~\cite{Gross:1973id} property of QCD to become a weak interaction at extremely high energy densities ( a property called 'asymptotic freedom'; we shall return to this later on), which could not hold together the bound hadrons. This idea turned out later to miss the real point of the QCD phase transition. But the first sketches of a QCD matter phase diagram, in terms of the variables temperature and density, had already been given~\cite{Kapusta:1979fh}. And, most importantly, there also appeared to be an experimental avenue to investigate hot, dense nuclear matter, perhaps up to, and beyond the hypothetical phase transition:
 
 Collisions of heavy nuclei (so-called 'Heavy Ion' collisions) at relativistic energy could compress and heat the initial nucleonic matter, up to, and beyond the critical energy density of QCD, thus estimated.
 
 In fact, concurrent experiments at the Berkeley Bevalac (a Linac injecting nuclei into the Bevatron synchrotron) had shown by 1980 that, at the prevailing modest energies in the low GeV per nucleon range, the two intersecting nuclear matter spheres stopped each other down creating a fireball at the center of mass coordinates. The major part of the initially longitudinal beam energy was trapped in it, leading to heating and compression. Of course we are not in cosmology here, nuclei are small and the fireball will disintegrate fast, so the crucial issue of thermal equilibrium attainment needed attention. Can we apply QCD thermodynamics, or hydrodynamics to the fireball? Phase transitions require a certain minimum ('relaxation') time to be completed. How fast does the fireball re-expand? Will thermodynamic processes leave a trace in the eventually emitted hadrons, or will these look like a trivial superposition of elementary nucleon collisions?
 
 The main question: could one define observable properties that should reveal characteristic stages or processes such as a typical plasma radiation, or characteristic changes of the initial 'partonic inventory' unambiguously due to cooking through a partonic plasma fireball? Toward the early 1980's some of these questions had found first affirmative answers, however tentative, from extrapolating the results obtained at the Bevalac were collective processes such as hydrodynamic flow of hadronic matter had been clearly observed. The most important step, however, came from the progress of thermal QCD theory solving the unsurmountable mathematical problems of low energy, 'non-perturbative' QCD numerically, in a lattice approximation. The existence of a phase change between hadrons and partons was shown, for the first time, with a plot by the US lattice group~\cite{Gottlieb:1987mq} showing the specific heat capacity of QCD matter as a function of the temperature. It showed a dramatic, steep upward jump at some critical temperature T$_c$. This signals the massive increase in the number of degrees of freedom, which should indeed take place when nucleons decompose into their constituent quarks at the phase boundary. Recall that a nucleon consists of three bound quarks which, moreover, each carry one out of three different colour charge units. They become the new set of degrees of freedom in a deconfined QCD plasma state. And, equally sensational, the critical temperature was found to be about 170 MeV (initially with considerable uncertainties), essentially equal to Hagedorns former upper boundary of matter composed of hadrons!
 
 \begin{figure}[htb]
 \centerline{\includegraphics [width=13cm] {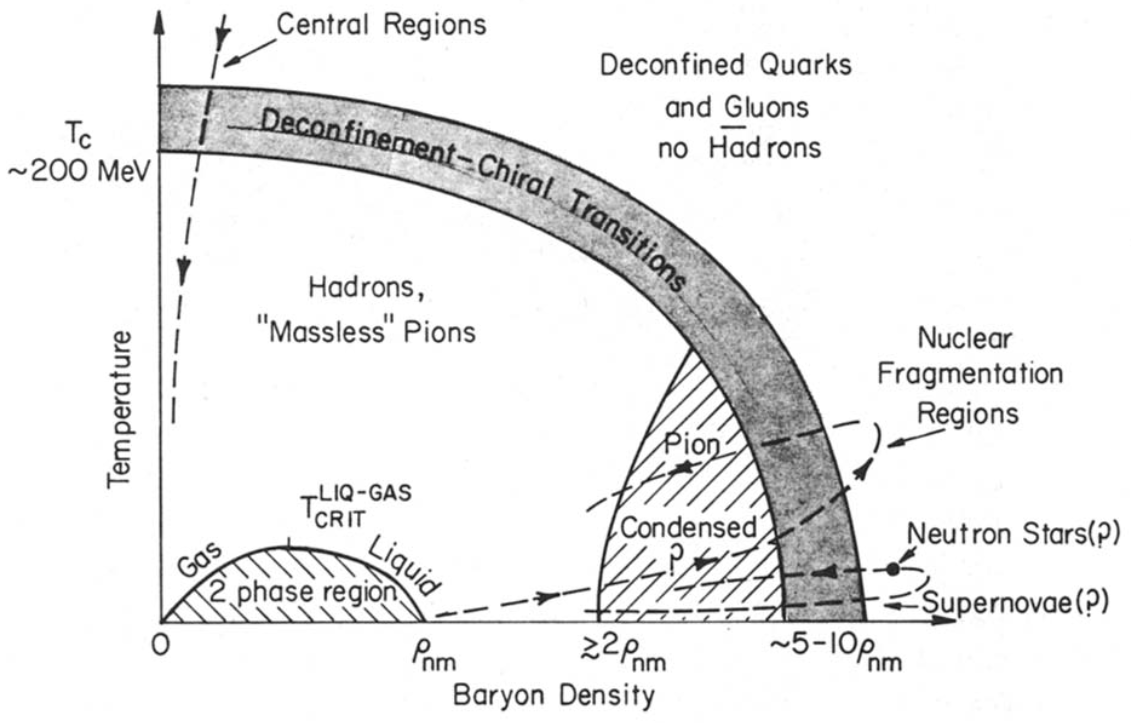}}
 \caption{An early sketch of the QCD phase diagram~\cite{Baym:1985tna}.}
 \label{Phasediagram}
 \end{figure}
 
 After these and other theoretical lessons learned, scientists got encouraged to sketch a universal phase diagram of strongly interacting matter~\cite{Baym:1985tna}, as shown in Figure~\ref{Phasediagram}. Looking at this plot one is reminded of the fact that a high energy density can be achieved, not only by heating but also by compressing. Both can, alone or in conjunction, drive QCD matter to the phase boundary. Thus the left hand part of the diagram refers to hot Big Bang dynamics and Quark-Gluon Plasma, the right hand side to gravitationally recompressed cold matter in supernovae, neutron stars, perhaps black holes. Are there quark stars?
 
 \section{Hot QCD matter research at CERN}
 At such heights of fascinating theory and speculation one MUST turn to experiments! Extrapolating from the Bevalac experiences it became clear that nuclear projectile energies in the range of 10 to 100 GeV per nucleon would be required in fixed target experiments at synchrotrons, in order to reach, and surpass the predicted QCD phase boundary. Clearly a task for synchrotron facilities such as the PS and SPS of CERN, and for the AGS at Brookhaven National Laboratory. First research proposals were thus directed at these laboratories in the period from 1982 to 1986. Concurrently, however, more far-reaching concepts were first formulated, aiming at nucleus-nucleus collisions at much higher center of mass energies that could be reached by colliders. Thus the idea was born to finish the construction of the temporarily abandoned superconducting ISABELLE collider project at Brookhaven~\cite{rhicproposal} which would reach up to ten times the center-of-mass (CM) energy per nucleon attainable at the CERN SPS. Also the Large Hadron Collider (LHC) at CERN appeared already on the horizon which, equipped with nuclear projectiles would reach up in energy by yet a further factor of 25. Unlimited future research opportunities were coming into view, and there occurred another phase transition as the number of interested scientists jumped up by more than an order of magnitude, from the modest Bevalac beginnings. And, in fact, ALL these projects did materialize in the following 30 years! Nuclear collision experiments started taking data in 1986 at the CERN SPS, in 1994 really heavy nuclear projectiles like Lead and Gold ($^{208}$Pb, $^{197}$Au) became available at the SPS, in 2001 at the newly built Relativistic Heavy Ion Collider, RHIC at Brookhaven, and in 2010 at the CERN LHC collider.
 
\begin{figure}[htb]
 \centerline{\includegraphics [width=1.0\textwidth] {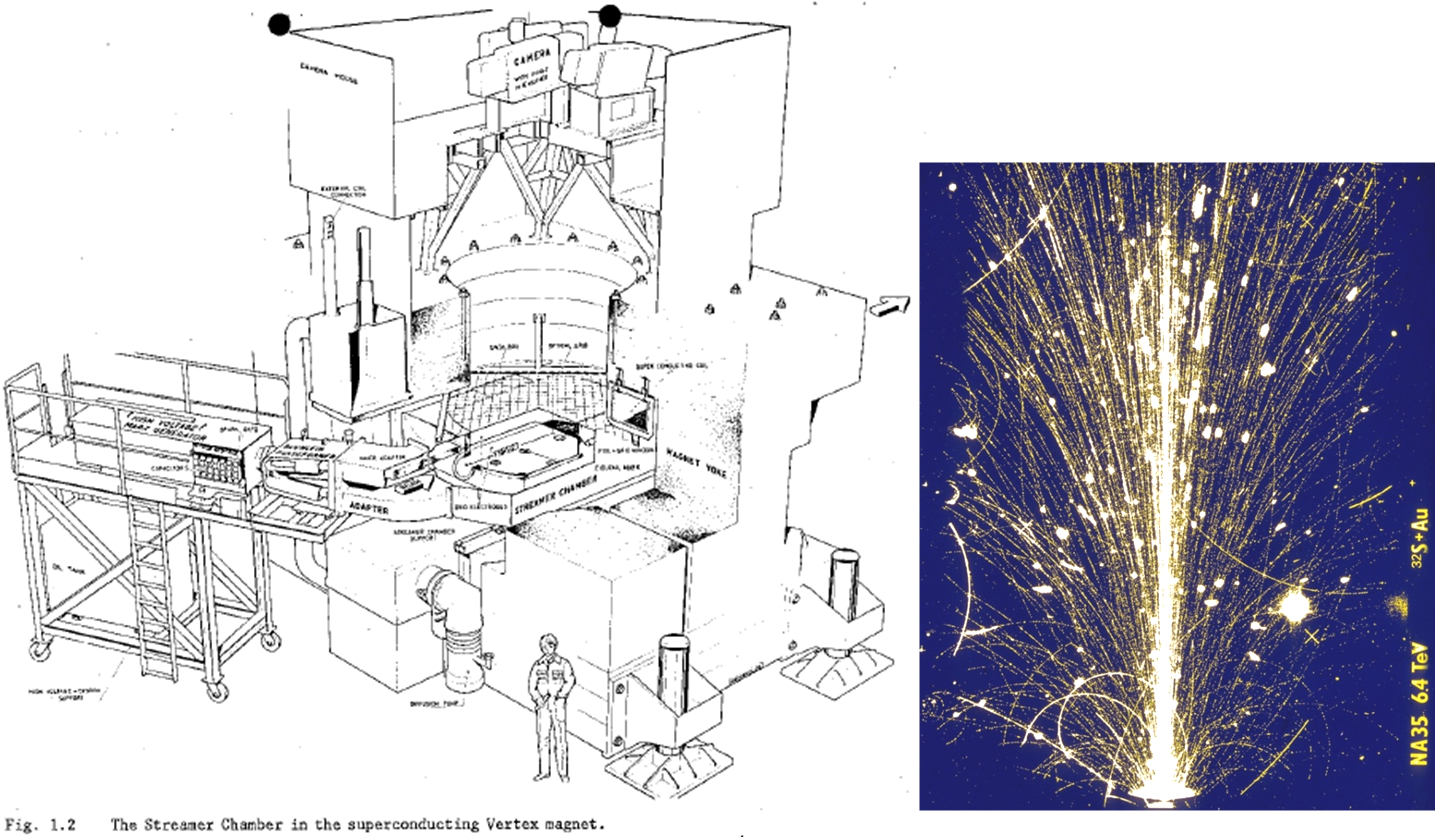}}
 \caption{NA35 experiment (left) and event display (right).}
 \label{SPSNa35}
 \end{figure}

\begin{figure}[htb]
 \centerline{\includegraphics [width=1.0\textwidth] {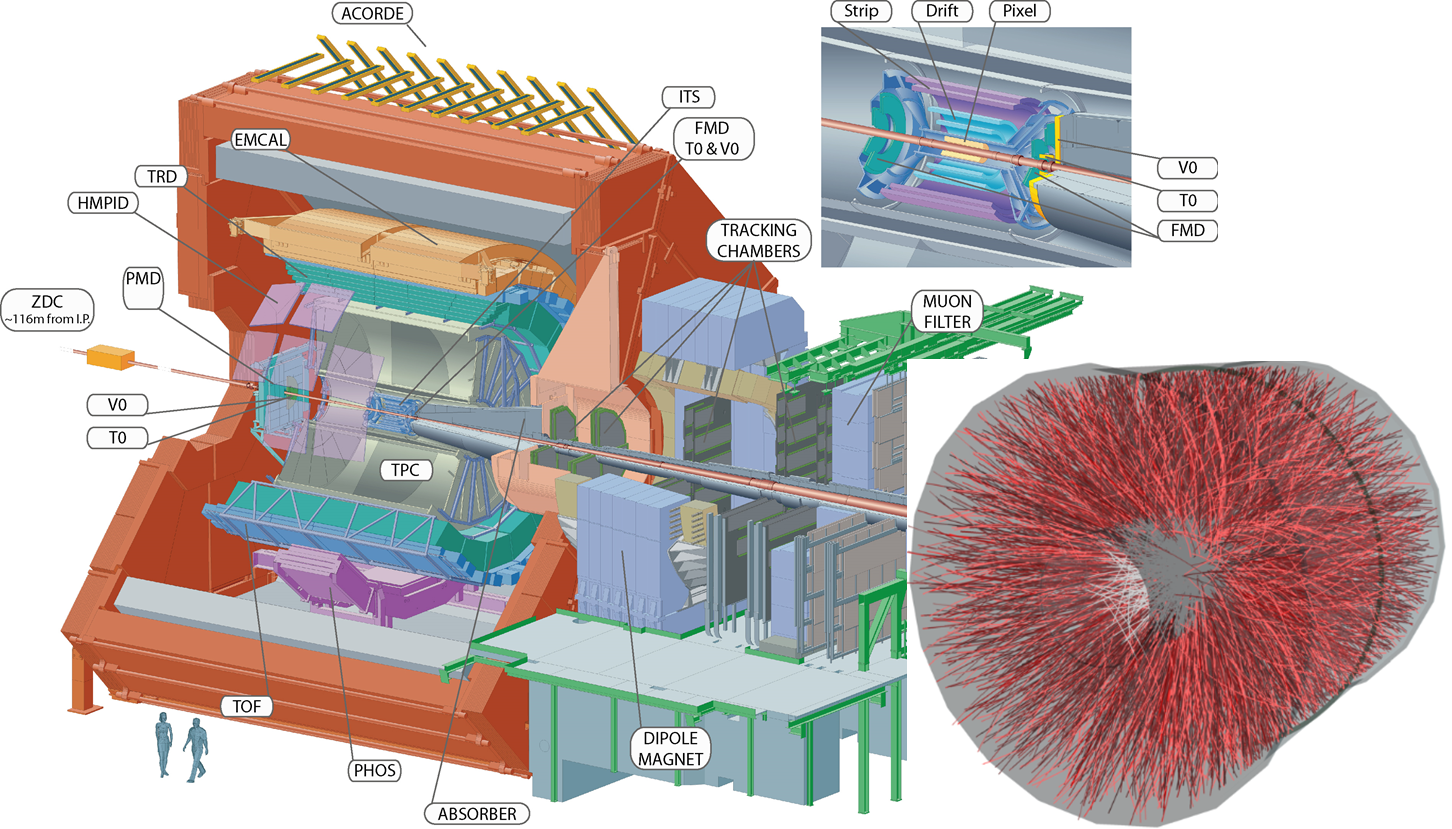}}
 \caption{Alice experiment (left) and event display (right).}
 \label{LHCAlice}
 \end{figure}

 Ever higher collisional energies and the accompanying relevant new physics observables have resulted in an overwhelming increase of experiment size, and complexity. We will turn below to a sketchy history of the corresponding physics ideas but wish, for now, to catch a typical glimpse at the developing dimensions of the experiments at CERN. To this end we show in Figure~\ref{SPSNa35} the layout of NA35, one of the first SPS experiments, and confront it in Figure~\ref{LHCAlice} with a sketch of the contemporary ALICE experiment at the LHC. The size of the international collaborations that have constructed these experiments, with significant help of CERN work force and funding, went up from about a hundred to more than a thousand physicists. Take the two experiments, illustrated above, for a typical example. NA35 was planned from 1981 onwards, and took SPS data from 1986 to 1992. It was constructed by about 80 physicists. The plans for the ALICE experiment~\cite{Fabjan:2011jb} at the LHC where first drafted in 1991, to be continually improved within the course of a long construction period that significantly extended upon the initially proposed instrumental techniques, to take first data at the onset of LHC experiments at the end of 2009. It united more than thousand physicists, and is planned to stay data taking for another decade, not to mention the to-be-expected final evaluation period. Such experiments require an unprecedented span of continuous preservation of instrumental expertise, information storage integrity, and stability, over decades, of the intellectual pursuit of the scientific goals. This latter aspect of 'big science' experiments is, perhaps, one of the most interesting sociological examples of human intellectual cooperation.

 The most immediate consequence of the increase in collisional center of mass energy, from 17.3 GeV at the SPS to (presently) 2.76 TeV at the LHC, is a rising number of created charged particles, from about 2'000 to about 20'000. These need to be resolved by tracking in a magnetic field, thus measuring their momenta, by track curvature, and their ionization intensity in the gas, along the track. The modern detector of choice is the Time Projection Chamber (TPC), employed in SPS NA49, in RHIC STAR, and LHC ALICE. Particle identification (there are pions, kaons, protons, electrons, etc.) is completed by an outer shell of detectors that measure each particles velocity. The tenfold rise in the number of charged particles produced in head-on collisions of Lead nuclei requires a larger tracking volume, and a much higher readout granularity, in the ALICE experiment. Moreover, the physics focus has shifted from SPS to LHC where so-called 'hard QCD processes', like jet production, came into the center of attention. The involved high momentum hadrons, from tens to hundreds of GeV/c, require higher tracking accuracy than the 'thermal' hadrons in the former SPS experiments. This resolution is proportional to $BL^2$ with B the magnetic field strength and L the tracking length inside the field. The incredible complexity of the tracking tasks is also illustrated in Figures~\ref{SPSNa35} and \ref{LHCAlice}. The former shows the track information in a single $^{32}$S+Au collision at the top SPS energy, 200 GeV per projectile nucleon, which is about 20 GeV per nucleon pair in the center of mass frame. The figure shows the tracks in the NA35 Streamer Chamber, a 3D photographically recorded gas detector. Its somewhat archaic tracking technique was replaced 1994 by the Time Projection Chamber (TPC) in the NA49 experiment succeeding NA35. The yet far higher tracking effort at ALICE (with 2.76 TeV per nucleon pair center of mass energy collisions) is illustrated in Figure~\ref{LHCAlice}; this TPC has over 600'000 electronic channels, producing a raw data flux of about 20 Gbyte per collision event. At 1 kHz event frequency the TPC yields 20 Tbytes per second of raw data, clearly at the upper end of today's digital electronics capacity.
 
 Let us note right here that a multiplicity of several thousand hadrons does not imply a dull accumulation of 'more of the same': there are subtle forms of global structures in each event, caused by a hydrodynamic collective expansion of partonic and hadronic QCD matter. This gives specific momentum kicks to all particles which are small individually but can be well quantified if all particles of an event are recorded, exhaustively. Furthermore, the very high multiplicity of charged particles, produced in a single event, results in a completely new phenomenon: the event originates from a single quantum mechanically coherent process but it can be individually analysed with statistical significance. Single events become self-analysing.
 
 \subsection{The acceleration of heavy nuclei at CERN}
 Synchrotrons normally accelerate protons or electrons, as well as their antiparticles. The protons have a charge to mass ratio of Q = 1. Stable nuclei have Q = 0.5 up to Calcium (charge 20, mass 40), but exhibit an excess of neutrons over protons from there on such that Q = 0.39 for $^{208}$Pb. The acceleration rate diminishes with lower Q; a proton synchrotron with 450 GeV top proton energy (the SPS) delivers Pb beams of up to 175 GeV per nucleon only, but stable acceleration modes can be accommodated. The problem is that the projectiles travel for about a million kilometres during acceleration, through the finite vacuum in the acceleration cavity. They will change their ionic charge state by stripping in the dilute gas, thus getting lost unless they are totally ionized already at injection, or if the vacuum is of extraordinary perfection. For proton acceleration this is of no concern, and thus synchrotrons tend to have a modest vacuum. They accept completely stripped ions only. The name 'Heavy Ion Physics' initially given to the field reflects the fact that fully stripped ions are needed, which can only be achieved in a complicated, multi-step pre-acceleration system. At its beginning a special ion source is required which delivers partially stripped ions out of an atomic plasma generated by strong electric and magnetic fields. It should give as high charge states as possible, and yet sufficient ionic current to be able to maintain beam stability in the subsequent synchrotrons, and to do experiments with sufficient event rate. During the early planning stage of CERN heavy ion research, 1982 to 1985, the preferred source type was the Electron Cyclotron Resonance (ECR) source~\cite{Geller:1981rz}. The first nuclear beams were Oxygen in 1986, a light nucleus, swiftly followed by Sulphur in 1988. A much more elaborate acceleration scheme was employed for Lead ions in the SPS that came into operation in 1994. With some modifications it is still used today at the LHC.
 
 A new high power ECR source produced Pb ions with charge states up to 20+. After charge state analysis this beam was accepted by the tanks of a newly built Linac which accelerated up to about 5 MeV per nucleon. Passing the beam through a thin stripper foil at the corresponding velocity produces a broad distribution of charge states around 50+ and the selection of a single charge state results in a 90\% loss of beam intensity at this stage. Now injecting into the first synchrotron, the so-called PS-Booster, one loses another large fraction of the precious beam because the synchrotron gets injected (at bottom field) for a few tens of microseconds only, and then again at the next acceleration cycle. The Booster vacuum was improved to $10^{-9}$ torr to minimize charge state changes. At the energy of 150 MeV per nucleon reached at extraction the Pb ion can still not be fully stripped, and a further grave intensity loss would result from spreading over several charge states. Thus the vacuum of the Proton Synchrotron (PS), the next element in line, had also to be improved to $10^{-9}$ torr, while still losing 50\% of the beam. The PS then extracts Pb(50+) at about 7 GeV per nucleon; stripping now produces fully ionized Pb(82+) nuclei for SPS injection. The PS has an acceleration cycle of one second duration whereas the SPS requires more than 15 seconds at top energy. The accelerator scientists thus can employ a complicated multi-turn injection technique at this stage, the SPS accommodating 4 successive PS extractions before its acceleration cavity is full of precisely positioned beam bunches; and then acceleration begins. The fourfold intensity gain far outweighs the slightly lengthened overall cycle duration. A final SPS energy of 158 GeV per nucleon was adopted for Lead projectiles. This outstanding facility could also accelerate all lighter elements compatible with the given ECR source technique, and operate over a wide range of energy, from about 15 to 160 GeV/A. Complete in 1994 it faced no competition worldwide until the turn-on of the Brookhaven RHIC collider, which offered about ten times higher energy in 2000.
 
 \subsection{The CERN SPS experiments and their physics}
 The first proposal~\cite{ Stock:1982aw} was submitted to CERN in 1982, by a GSI-LBL-Heidelberg- Marburg-Warsaw Collaboration of research groups established in nuclear physics, and concurrent Bevalac or Dubna Synchrophasotron engagement. It called for the establishment of an extracted CERN PS beam of Oxygen where GSI would purchase an ECR source from the Grenoble group of R. Geller, and LBL would construct an RFQ micro-Linac, then to inject the existing Linac1 of CERN, followed by Booster and PS, extracting at 13 GeV/A. It was proposed to perform two parallel experiments, based on experience with two concurrent Bevalac experiments, the multi-segmented scintillator Plastic Ball~\cite{Baden:1982pw} and the visual tracking Streamer Chamber spectrometer~\cite{Sandoval:1982cn}. The former would investigate hydrodynamic nuclear matter flow, the latter meson production with a look at phase transition signals.
 
 This initial proposal was accepted by CERN; however the accommodation of further experiments in the East Hall PS extraction area met with substantial difficulty. Of seminal consequence was then the suggestion~\cite{Klapisch:1985ry} of CERN management to transport the PS beam to the SPS and distribute the resulting beams, at 200 GeV/A, via the external SPS beam line system to the then little used, huge experimental halls in the North and West of the SPS, where the former SPS proton beam experiments had been conducted.
This idea catalysed the much more forward-looking idea of a full-fledged SPS heavy ion acceleration program, with beam energies ranging up to 200 GeV/A, which was enthusiastically welcomed as it also met with the intentions of several CERN experiment groups to establish a continuation of the formerly abandoned initiatives toward heavy ion experiments at the CERN ISR collider.
A wealth of established experimental infrastructure was available here and, most importantly, three intact experiments, with still existing physics collaborations, as well as technicians: the huge magnetic hadron spectrometer OMEGA, the dilepton spectrometer from experiment NA10, and an almost complete Streamer Chamber plus calorimeter experiment, with a 400 ton superconducting dipole magnet, from experiments NA5 and NA24. The turn to the SPS really was a stroke of genius!
 
 The anticipated SPS research program of CERN attracted further groups from nuclear physics but also a fraction of the particle physics groups working at CERN already. The Omega spectrometer group reshaped as WA85, the dilepton spectrometer as NA38, the Streamer Chamber experiment as NA35, the large calorimeter experiment NA34 (many of whose members were formerly engaged in a CERN ISR study of $^4$He collisions - a precursor of the SPS program~\cite{Fischer:1980gb}) became NA34-2. The Plastic Ball spectrometer moved from LBL and was amended with Lead glass electromagnetic calorimetry, to become WA80. Initially, only experiment NA45 was completely newly constructed, a double Cherenkov (RICH) magnetic spectrometer for dielectron spectroscopy. The culminating part of the program was carried out from 1994 to 2002, with Lead nuclei, also including lower energies, 20, 30, 40 and 80 GeV/A. This setup was reactivated in 2005 with Indium ($^{115}$In) beams for NA60, a high precision charmonium and di-muon spectrometer constructed on the basis of former NA38 and NA50, and with Lead beams for NA61, a large acceptance hadron spectrometer based on NA49. We list below the experiments from the main Lead beam program:
 
 NA44 Small angle focusing magnetic spectrometer for antiprotons and kaons.

 NA45 Double Cherenkov ring imaging magnetic spectrometer for di-electrons.

 NA49 Large acceptance TPC and calorimeter spectrometer for all hadrons.

 NA50 Magnetic di-muon spectrometer, EM calorimeter, for vector Mesons.

 NA52 Beam line spectrometer looking for strangelets.

 WA97/NA57 Hyperon and antihyperon spectrometer with Si pixel technique, ex WA85.

 WA98 Large acceptance hadron and photon spectrometer for direct photons.

 NA60 and NA61 followed later, as stated above.
 
 \section{Results at the Millenium}
 We shall briefly sketch below a (subjective) selection of seven physics observables, emerging from the SPS program and representing the increasing understanding of QCD matter and QCD phases. In this section we have to ask the reader for some patience because we have to turn to a more detailed physics argumentation. The experiments addressed mostly 'to be or not to be' questions. A quantitative description of the QCD plasma has resulted from the next following research era, at the colliders RHIC at Brookhaven and LHC at CERN, as we will show at the end of this article. Now let us take a closer look at the main topics of SPS research.
 
 \subsection{Fireball energy density}
 Calorimeter experiments (NA34, NA49, WA98) measured the total transversal energy produced in head-on collisions of two $^{208}$Pb nuclei ~\cite{Margetis:1994tt}. Central, head-on collisions fall into the tail of the distribution. Here one has, on average, about 190 participating nucleon pairs from the initial target-projectile nuclear density distributions. So we know the collision geometry, the total initial energy in the CM system, and the newly created transversal energy. A formula derived by Bjorken~\cite{Bjorken:1982qr} provides for an estimate of the corresponding energy density in the primordial fireball volume. In the present case it results in 3.0+/-0.6 GeV per cubic Fermi. Comparing to a 'year 2000' view of the parton-hadron phase transition from Lattice QCD~\cite{Hands:2001ve} one finds that we are just above the phase transition at this energy density, and that the critical QCD energy density is about 1 GeV per cubic Fermi
 
 \subsection{Fireball temperature}
 Lattice QCD also gives a plasma temperature estimate, prevailing at 3 GeV/fm$^3$ energy density, of T = 210 MeV. The photons from thermal plasma radiation should escape from the fireball unaffected because they lack strong interaction. A first measurement~\cite{Aggarwal:2000th} was undertaken by WA98 of the so-called 'direct photons', which result from a meticulous subtraction of the trivial photon fraction resulting from electromagnetic decay of neutral mesons, that occurs much later. These data can be described\cite{Srivastava:2000pv} with a plasma temperature in the 200 -250 MeV domain, in accord with the Lattice QCD temperature estimate given above.
 
 \subsection{Hadrons form at T = 160 $\pm$ 10 MeV: close to lattice QCD prediction}

\begin{figure}[htb]
\begin{tabular}{cc}
\begin{minipage}{.55\textwidth}
\centerline{\includegraphics[width=1.0\textwidth]{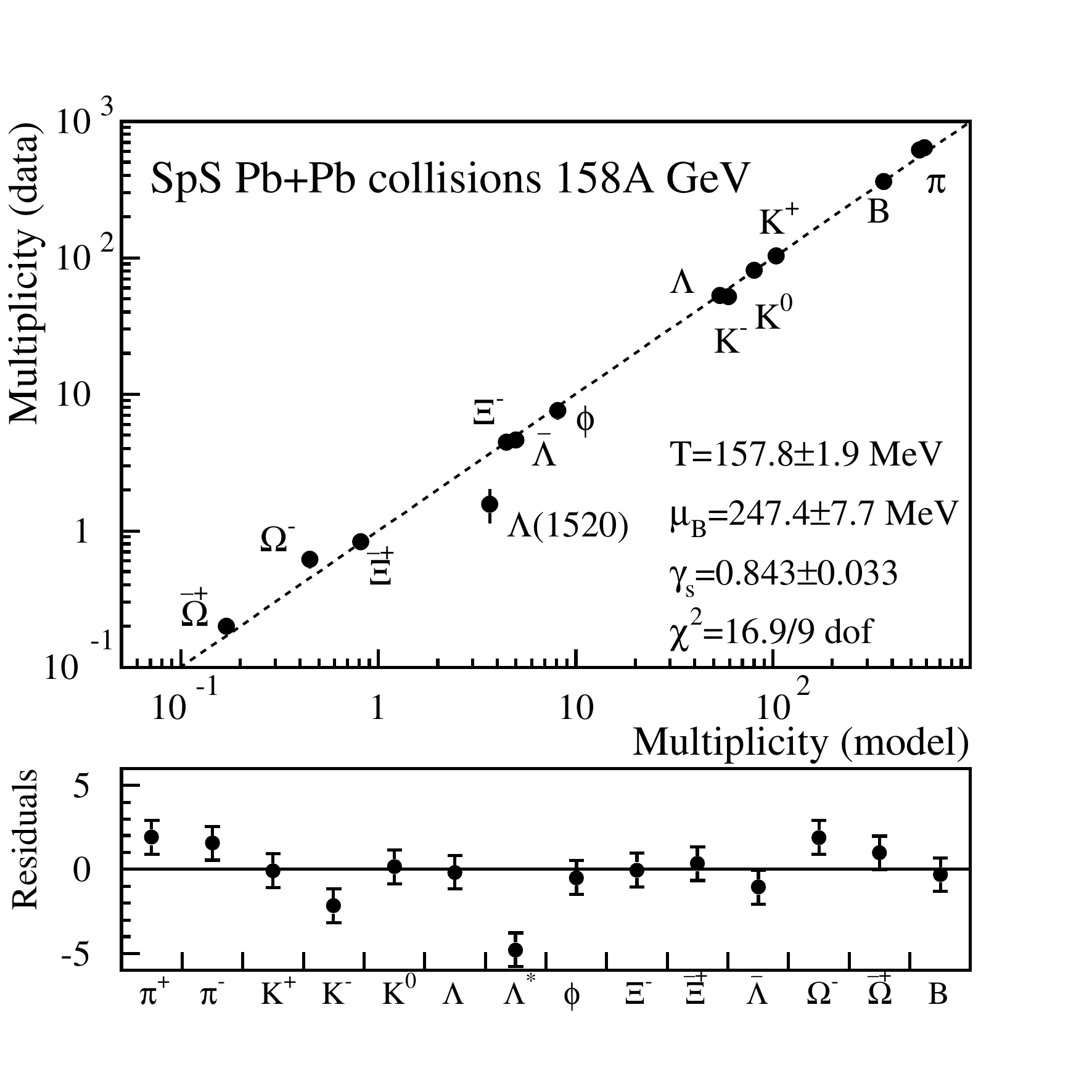}} 
\caption{Hadron multiplicities vs. prediction of the Statistical Model.}
\label{LHCRSFig11}
\end{minipage} & \begin{minipage}{.41\textwidth}
\centerline{\includegraphics[width=1.0\textwidth]{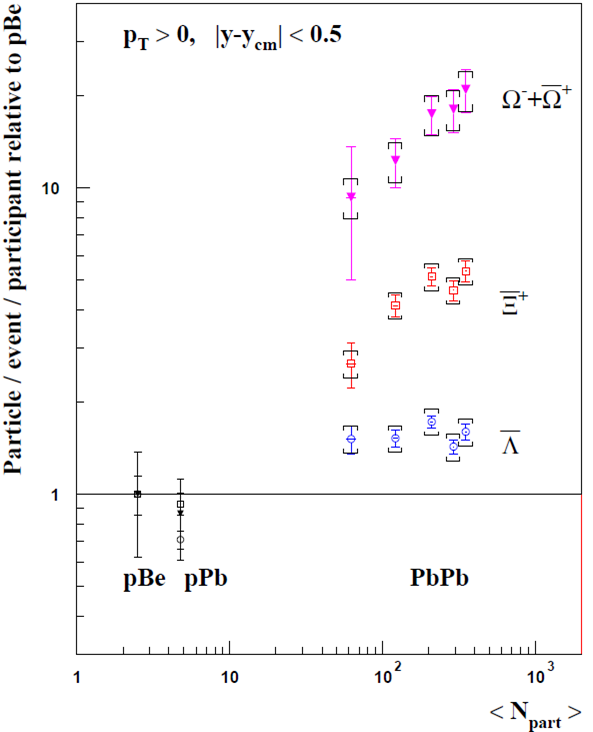}}
\caption{Strangeness enhancement in Pb+Pb Hyperon yields.}
\label{LHCRSFig12}
\end{minipage}
\end{tabular}
\end{figure}

 If a primordial QCD parton plasma is formed in central Pb+Pb collisions at top SPS energy, subsequent expansive cooling will bring the fireball volume back down to the QCD parton-hadron phase boundary, where confinement enforces hadron formation. If a critical QCD hadronisation temperature uniformly governs the fireball volume the various hadronic species will be simultaneously produced in proportion to their so-called statistical weights, a universal law that was discovered by E. Fermi and carries the name 'Fermi's golden rule'. This is articulated in the Statistical Hadronisation Model (SHM) which predicts a universal yield order among the produced hadrons, with just only one essential parameter: the temperature T prevailing at birth of the hadronic final state~\cite{BraunMunzinger:2003zd}. Hadron multiplicity distributions were systematically measured by NA44, NA49 and WA 57. Figure~\ref{LHCRSFig11} shows an example~\cite{Becattini:2003wp} of a SHM fit to hadron multiplicities per collision event, observed at top SPS energy by NA49 in central Pb+Pb collisions. It gives a hadronisation temperature of T = 158 $\pm$ 5 MeV, in close agreement with the critical temperature T$_c$ predicted by Lattice QCD. At this low temperature we are very far away from QCD asymptotic freedom~\cite{Collins:1974ky, Gross:1973id}. The transition from deconfined to confined QCD matter must be driven by other, genuinely non-perturbative QCD mechanisms.
 
 
 \subsection{Strange baryon and antibaryon production is enhanced}
 The 'cooking' in an extended, thermal QCD matter state enhances the production rate of strange hadrons, in particular of the strangeness 1, 2 and 3 carrying so-called hyperons (Lambda, Xi and Omega, as well as their antiparticles). This observation, made in nucleus-nucleus collisions, was called strangeness enhancement.

 Figure~\ref{LHCRSFig12} shows a measurement from NA57~\cite{Dainese:2005vk}. Pb+Pb collisions at top SPS energy are considered here, and the various hyperon and antihyperon multiplicities are observed in different, successive windows of collisional impact geometry, as monitored by the (simultaneously determined) number of nucleons participating in the collision, that increases toward central, head-on collision geometry. The fireball volume is growing with N(part). The yields are shown here as multiplicity per participant nucleon number, N(part). Moreover, they are normalized relative to the yields per participant observed in proton-nucleus collisions where no QCD plasma is expected to form. If the fireball outcome was merely a trivial superposition of elementary nucleon-nucleus collision multiplicities (here represented by p+Be and p+Pb results at similar energy) the hyperon production rate should simply stay constant. However, on the contrary, we observe an increase of the relative yields with Pb+Pb centrality, by up to above a factor of 10, for Omega and Antiomega hyperons. Pb+Pb collisions are not a trivial superposition of p+A collisions, and strangeness enhancement grows with increasing fireball volume, but saturates toward central collisions. The fireball volume is seen to act collectively, as implied by the grand canonical (large volume limit) Statistical Hadronisation Model ( Figure 4).
 
 \subsection{Charmonium (J/Psi) suppression reveals QCD plasma formation}
 
\begin{figure}[htb]
\begin{tabular}{cc}
\begin{minipage}{.54\textwidth}
\centerline{\includegraphics[width=1.0\textwidth]{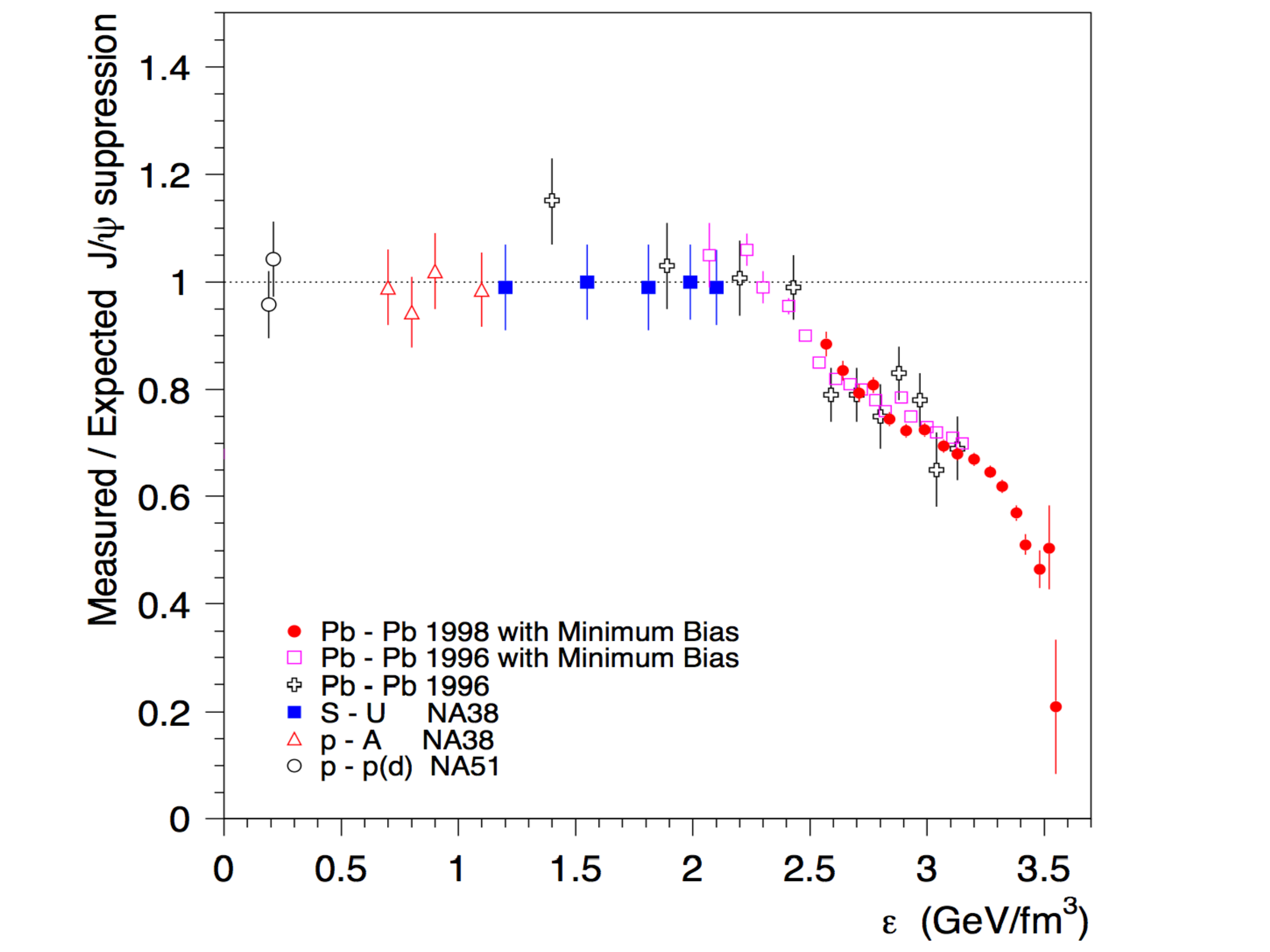}} 
\caption{J/Psi suppression vs. centrality at the SPS.}
\label{LHCRSFig13}
\end{minipage} & \begin{minipage}{.42\textwidth}
\centerline{\includegraphics[width=1.0\textwidth]{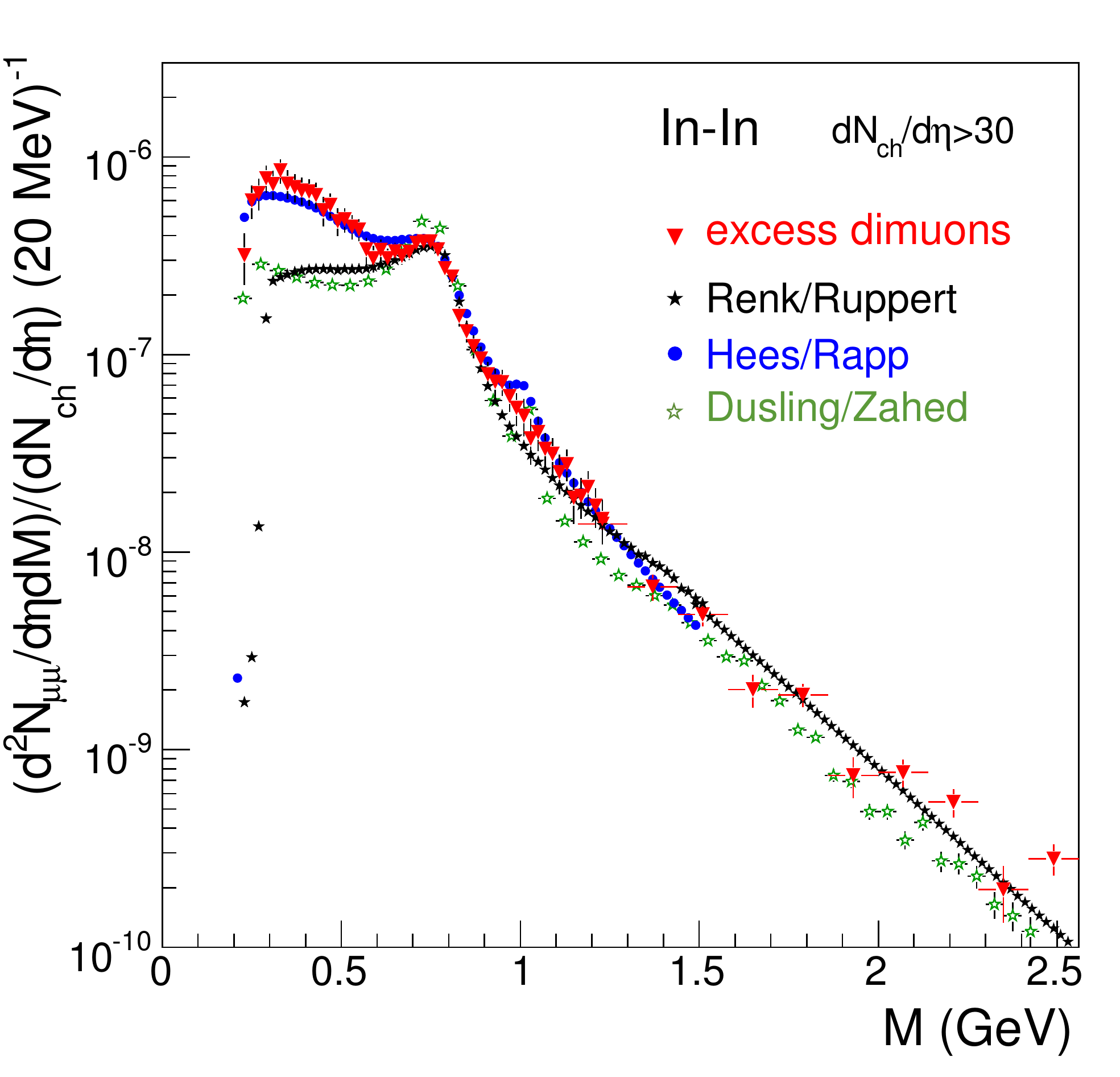}}
\caption{Di-muon invariant mass spectrum of the excess yield in In+In.}
\label{LHCRSFig14}
\end{minipage}
\end{tabular}
\end{figure}

 J/Psi vector mesons are called charmonia. They arise from c-cbar, charm-anticharm quark pairs that are produced by 'hard' parton collisions in the initial phase of an A+A collision. If, subsequently, a deconfined QCD plasma phase governs the dynamical evolution, the primordial c-cbar pairs co-travel with the medium and can break up by QCD Debye screening of the interquark colour exchange force~\cite{Matsui:1986dk}, instead of evolving into the final charmonium states. As a result, J/Psi production in A+A collisions will be suppressed by QCD plasma formation. Figure~\ref{LHCRSFig13} shows the experimental verification by experiments NA38 and NA50 (as well as later by NA60). J/Psi production is illustrated for a number of different reactions at various centralities~\cite{ Abreu:2000ni}. The data can be represented on a common scale, the fireball energy density introduced previously, which increases monotonically with increasing number of target and projectile participants. We would now go on simply plotting the respective yields per participant, to find a suppression uniformly increasing with energy density: J/Psi production gets suppressed toward central collisions, the deconfinement signal we wanted to quantify. However, there also occurs absorption of c-cbar pairs in ordinary, cold nuclear matter. It gets quantified by measuring p+A collisions at similar energy, and a complicated method is employed to simulate the effect of this 'normal' absorption cross section in an A+A collision geometry (as if it were cold). Figure~\ref{LHCRSFig13} then, finally, shows the real J/Psi relative to a constructed yield based on the normal cold absorption only. One calls deviations from unity in this ratio the 'anomalous suppression of J/Psi'. Figure~\ref{LHCRSFig13} shows that this suppression gets stronger with increasing energy density: an ever smaller fraction of the initially produced c-cbar pairs find their way to become final hadrons because they dissolve in the plasma.

 \subsection{QCD chiral symmetry restoration: hadrons melt near T$_c$}
 What becomes of the large hadron mass, and of their quark 'wave functions', once hadronic matter approaches the QCD phase boundary? The plasma quarks are massless in QCD, their small masses, in the low MeV range for the up and down quarks, out of which hadronic matter consists, stem from the Higgs mechanism. QCD has the property of chiral symmetry, a peculiarity of massless particles propagating at the speed of light: if their spin turns them left or right handed with respect to the direction of their momentum they will stay that way forever. This symmetry is completely lost ('broken') in hadrons where quarks dress up with massive vacuum polarization clouds. This transition can be observed. Does it coincide with the QCD deconfinement process occurring at the critical temperature T$_c$? I.e. will we see 'dissolving' hadron wave functions near T$_c$? All neutral, non-strange mesons decay to electron and/or muon pairs. Their intensity and invariant mass should reveal a melting of the mesonic wave functions. The spectroscopy of lepton pairs was pioneered by NA45 and perfectioned (for muon pairs) by NA60. One constructs an 'invariant mass' spectrum from the observed momentum spectra of both leptons, in which each meson species creates a peak at its proper rest mass, of characteristic height, and width. A simple superposition of the known, in vacuum di-lepton decays (called the 'cocktail') should fail to reproduce the observed invariant mass spectrum if the fraction of overall dilepton decays of mesons that is created in the immediate vicinity of the deconfinement phase transition temperature exhibits effects of the concurrent chiral phase transition of QCD. We show the excess of the data over the cocktail prediction in the $\mu^+ \mu^-$ invariant mass spectrum obtained by NA60~\cite{Damjanovic:2005ni,Specht:2010xu} in Figure 7. It was measured in In+In collisions at top SPS energy. The excess yield up to about 1 GeV is attributed to emission from the immediate vicinity of T$_c$. There it is dominated by an in-medium process which is strongly enhanced at the high matter density, prevailing here: two pion annihilation $\pi^+ + \pi^-$ to an intermediate rho vector meson that decays to $\mu^+ \mu^-$. The rho turns out to be very strongly broadened in the fireball medium near T$_c$, thus accounting for the excess of the data over the cocktail expectation: a first indication of QCD chiral symmetry restoration occurring at the phase boundary to partons. Furthermore, the exponential tail in Fig.7, upward from about 1 GeV, reveals the contribution of quark-antiquark annihilation to lepton pairs, from the preceding plasma phase. The Planck-like spectrum reflects an average plasma temperature of about 220
MeV~\cite{Specht:2010xu}: the first "direct" plasma signal at SPS energy.

 \subsection{The fireball matter exhibits collective hydrodynamic flow}
 Consider the very initial stage of a Pb+Pb collision, not head-on but at a finite impact distance between the trajectories of the impinging nuclei (called a semi-peripheral collision). Looking along the incoming projectile we see the overlapping areas of the target and projectile density distributions, as projected onto the plane transverse to the projectile direction. The projectile will carve out an ellipsoidal sector from the target, and a correspondingly shaped fireball will develop next. In it the created energy density falls down faster along the impact vector direction ( in the so-called reaction plane) than perpendicular to it. Thus the expansion pressure is higher in the reaction plane, leading to higher expansion momenta of matter emitted in-plane, as compared to the out of plane direction. As a consequence, the initially partonic, and subsequently hadronic expansion pattern will acquire a spatial anisotropy. If a collective, hydrodynamic outward flow of matter sets in at this primordial time it will preserve this particular anisotropy signal. From its origin in an elliptically shaped fireball it is called 'elliptic flow'. If the in-medium viscosity is small the initial anisotropic expansion pattern is carried on to a finally observed hadronic emission anisotropy, as quantified by the second harmonic coefficient $v_2$ of a Fourier decomposition of the final, collective hadronic emission pattern. Its measurement by NA49~\cite{Alt:2003ab} observed both pion and proton $v_2$ in semi-peripheral Pb+Pb collisions at top SPS energy. Fireball partonic matter flows collectively, in its expansion, much like a liquid, to be described by QCD hydrodynamics. And, moreover, this liquid appears to have rather little dissipative viscosity, or else the primordial pressure anisotropy would not survive in hadrons, emitted much later, after fireball expansion. A topic of very high significance as became obvious in the course of later, more detailed investigations at RHIC and LHC. We shall return to it below.

 
 \subsection{Summary of SPS results and interlude at RHIC}

 An appraisal of the SPS program was made just prior to RHIC turn on, in 2000~\cite{Heinz:2000bk}, based on a 'common assessment' of the results collected and published over the preceding years. It concluded that \emph{ 'compelling evidence has been found for a new state of matter, featuring many of the characteristics expected for a Quark-Gluon Plasma'}~\cite{ CernPress}. This conclusion was based primarily on three of the experimental observations mentioned before: the copious production of hadrons containing strange quarks ('strangeness enhancement'), the reduced production of $J/\Psi$ mesons ('anomalous $J/\Psi$ suppression'), and the yields of low mass lepton pairs ('rho melting'). These three signals most closely resembled the predicted hallmarks of the QGP, namely thermalisation, deconfinement, and chiral symmetry restoration.
 
 The experimental results have all stood the test of time and have been amply confirmed and refined in the years thereafter. The essence of the assessment however, seems in hindsight today more compelling than in 2000, given for example much improved low mass lepton pair results from the SPS NA60 experiment~\cite{Specht:2010xu}, which started taking data only in 2005, and new results and insights gained from the RHIC and LHC programs which are described below.

 In the summer of 2000, the core of heavy ion activity shifted from CERN to BNL, when the dedicated Relativistic Heavy Ion Collider started operation with Au+Au collisions at 130 GeV center of mass energy. Reaching its design energy of 200 GeV the year after, RHIC stayed at the energy frontier for the next decade until the arrival of LHC. While a detailed description of RHICs scientific legacy is beyond the scope of this article, a short summary of the main highlights is given below to set the stage for the next chapter at CERN with the LHC.
 
 The initial results from RHIC were summarized and assessed in 2005, based on a comprehensive (re)analysis of the first few years of RHIC running~\cite{rhichwhitepaper}. The experiments concluded that at RHIC \emph{'a new state of hot, dense matter'} was created \emph{'out of the quarks and gluons ... but it is a state quite different and even more remarkable than had been predicted'}~\cite{BnlPress}. Unlike the expectation, with hindsight overly naive, that the QGP would resemble an almost ideal gas of weakly coupled quarks and gluons, the hot matter was found to behave like an extremely strongly interacting, almost perfect liquid, sometimes called the sQGP (where the 's' stand for 'strongly interacting'). It is almost opaque and absorbs much of the energy of any fast parton which travels through -- a process referred to as 'jet quenching' -- and it reacts to pressure gradients by flowing almost unimpeded and with very little internal friction (i.e. has very small shear viscosity) ~\cite{Muller:2006ee}. The shear viscosity over entropy ratio, $\eta/s$ was found to be compatible with a conjectured lower bound of $ \eta/s \ge 1/4\pi $ ($\hbar = k_B = 1$), a limiting value reached in a very strongly interacting system when the mean free path approaches the quantum limit, the Compton wavelength.
 
 Also at RHIC, the crucial experimental results as well as the inferred characteristics of the QGP --- a 'hot, strongly interacting, nearly perfect liquid' -- stood the test of time~\cite{Muller:2012zq}.

 \section{Heavy Ion Physics at the LHC}
 Prior to LHC, 25 years of heavy ion experimentation had already revealed a \textbf{'QGP-like'} state at the SPS and not '\textbf{the} QGP', but '\textbf{a} sQGP' at RHIC. With the discovery phase considered to be essentially over~\cite{Schukraft:2011na}, a main goal for the heavy ion program at LHC was increased precision to better characterise this new state of matter, making use of the particular strength of the LHC, i.e. the huge increase in beam energy and a powerful new generation of large acceptance state-of-the-art experiments. These include the dedicated heavy ion detector ALICE~\cite{Fabjan:2011jb}, as well as the general purpose pp experiments ATLAS and CMS, which both participate fully in the heavy ion program, and finally LHCb, which joins in for p-nucleus collisions. The larger cross section for hard probes and the higher particle density at LHC creates a QGP which should be 'hotter, larger, and longer living'. And indeed, LHC made significant progress towards increasing the precision on shear viscosity (section~\ref{eflow}) and plasma opacity (section~\ref{jquench}) already during the first two years of ion running~\cite{Muller:2012zq,JSnobelsymp}. However, when dealing with QCD in the nonpertubative regime, surprises should not come as a surprise, and a number of unexpected findings at LHC have helped shed new light on some old problems or issues, for example on particle production (section~\ref{pratio}) and $J/\Psi$ suppression(section~\ref{jpsi}). And finally the very first discovery made at LHC was the appearance of a mysterious long range 'ridge' correlation in high multiplicity $pp$ reactions (section~\ref{ridge}). It reappeared later -- and much stronger -- in the 2012 p-nucleus run, making it of great interest, and presumably of great relevance, to hot and dense matter physics, even if it's ultimate cause and connection to similar phenomena in nuclear collisions is as of today not finally settled.
 
 \subsection{Hadron Formation} \label{pratio}
 
 Measuring identified particles at LHC was considered a somewhat boring but necessary exercise, as finding thermal particle ratios essentially identical to the ones measured at SPS and RHIC (save expected differences related to the ratios of particles and antiparticles) was thought to be one of the safest predictions~\cite{Abreu:2007kv}. It therefore came as a surprise when some particle fractions, in particular for the mundane proton, one of the most frequently produced hadrons, were found to differ considerably from expectations (and, to a lesser extent, from the ones measured at RHIC), while others, including those for multi-strange hyperons, were well in line with thermal predictions.
 
 Possible reasons being discussed range from mere adjustments of thermal model Parameters, over the consideration of hitherto neglected final state interactions and sequential freeze-out of individual hadron species, to the consideration of different transition temperatures for different quark flavours.
 The final resolution of the 'proton puzzle' is still outstanding and will probably require a more complete set of particle ratio measurements at LHC as well as revisiting the RHIC results to confirm with better significance if particle ratios in central nuclear collisions indeed evolve with energy. Whichever explanation will finally prevail, the unexpected LHC results are a welcome fresh input likely to advance our understanding of the remarkable success of the statistical model of hadron production.
 
 \subsection{Elliptic flow} \label{eflow}
 The observation of robust collective flow phenomena in heavy ion reactions at fixed target energies and at RHIC is the most direct evidence for the creation of a strongly interacting, macroscopic (i.e. large compared to the mean free path) and dense matter system in nuclear collisions. Analysed in terms of a Fourier expansion of the azimuthal charged particle density $dN_{ch}/d\phi$ with respect to the reaction plane ($\phi=0$), the first order component ($v_1 \propto \cos(\phi)$) is called directed and the second order component ($v_2 \propto \cos(2\phi)$) is called elliptic flow (recall sect.4.7). Matter properties like the equation of state, sound velocity or shear viscosity, can be extracted by comparing measurements and hydrodynamic model calculations of elliptic (i.e. azimuth dependent) and radial (azimuthally averaged) flow. Flow however depends not only on the properties of the hydrodynamic evolution but also on initial conditions, in particular the geometrical distribution of energy density within the primordial nuclear overlap zone. The resulting pressure gradients should thus reveal, in particular, possible effects of gluon saturation in the initial stage, as postulated in the Colour Glass Condensate (CGC) model.
 
 When first azimuthal flow data from the LHC became available in early 2011, the evidence from all three experiments, as well as new results shown by the two RHIC collaborations, was overwhelming~\cite{Schutz:2011zz}: The collective flow patterns in heavy ion collisions were much more complex with measurable and significant Fourier coefficients up to at least 6th order ($v_1, v_2, ..v_6$)! Today these patterns are understood to arise from fluctuations, event-by-event, of the initial geometry (i.e. pressure gradients) caused by the stochastic nature of nucleon-nucleon collisions and/or by a CGC initial state.
 
 The complex correlation patterns had actually been strong and clearly visible since many years; however, before 2011, they were in general not recognized as hydrodynamic in origin but discussed in terms of fancy names ('near side ridge, away side cone') and fancy explanations ('gluon Cerenkov radiation, Mach cone, ..')~\cite{Nagle:2009wr}. At LHC, the large acceptance of the experiments, together with the high particle density (as a collective effect, the flow signal increases strongly with multiplicity) made the observation and interpretation straightforward and unambiguous.
 
 The fact that energy density fluctuations on the scale of a fraction of the nuclear radius in the initial state are faithfully converted into measurable velocity fluctuations in the final state was a most amazing, and also most useful, discovery: One could not only identify the on average almond shaped collision zone, but recognize much finer structures of individual nuclear collisions. The analysis of flow has been invigorated and is advancing rapidly ever since~\cite{Heinz:2013th}, with direct measurements of the fluctuation spectrum~\cite{Aad:2013xma}, using event-by-event measurement and selection of flow as an analysis tool~\cite{Schukraft:2012ah}, and even finding non-linear mode mixing between different harmonics~\cite{Qiu:2012uy}. Like temperature fluctuations in the cosmic microwave background radiation, which can be mapped to initial state density fluctuations in the early Universe, collective flow fluctuations strongly constrain the initial conditions and therefore allows a better measurement of fluid properties. Since 2011, the limit for the shear viscosity has come down by a factor of two ($\eta/s < (2-3) x 1/4\pi$) and is now precise enough to even see a hint of a temperature dependence, slightly increasing from RHIC to LHC~\cite{Muller:2013laa}. Future improvements in data accuracy and hydro modelling should either further improve the limit, or give a finite value for $\eta/s$. In either case, improved precision is relevant as the shear viscosity is directly related to the in-medium cross section and therefore contains information about the degrees of freedom relevant in the sQGP via the strength and temperature dependence of their interactions.
 
 \subsection{Jet quenching} \label{jquench}

 High energy partons interact with the medium and lose energy, primarily through induced gluon radiation and, to a smaller extent, elastic scattering~\cite{Majumder:2010qh}. The amount of energy lost, $\Delta E$, is expected to depend on medium properties, in particular the opacity and the path length L inside the medium, with different models predicting a linear (elastic $\Delta E $), quadratic (radiative $\Delta E$), and even cubic (AdS/CFT) dependence on L. In addition, $\Delta E$ also depends on the parton type via the colour charge (quark versus gluon), the parton mass via formation time and interference effects (light versus heavy quarks), and finally somewhat on the jet energy. The total jet energy is of course conserved and the energy lost by the leading parton appears mostly in radiated gluons, leading in effect to a modified softer fragmentation function.
 Jet quenching (i.e. measuring the modified fragmentation functions) is therefore a very rich observable which probes not only properties of the medium but also properties of the strong interaction.

\begin{figure}[htb]
\begin{tabular}{cc}
\begin{minipage}{.44\textwidth}
\centerline{\includegraphics[width=1.0\textwidth]{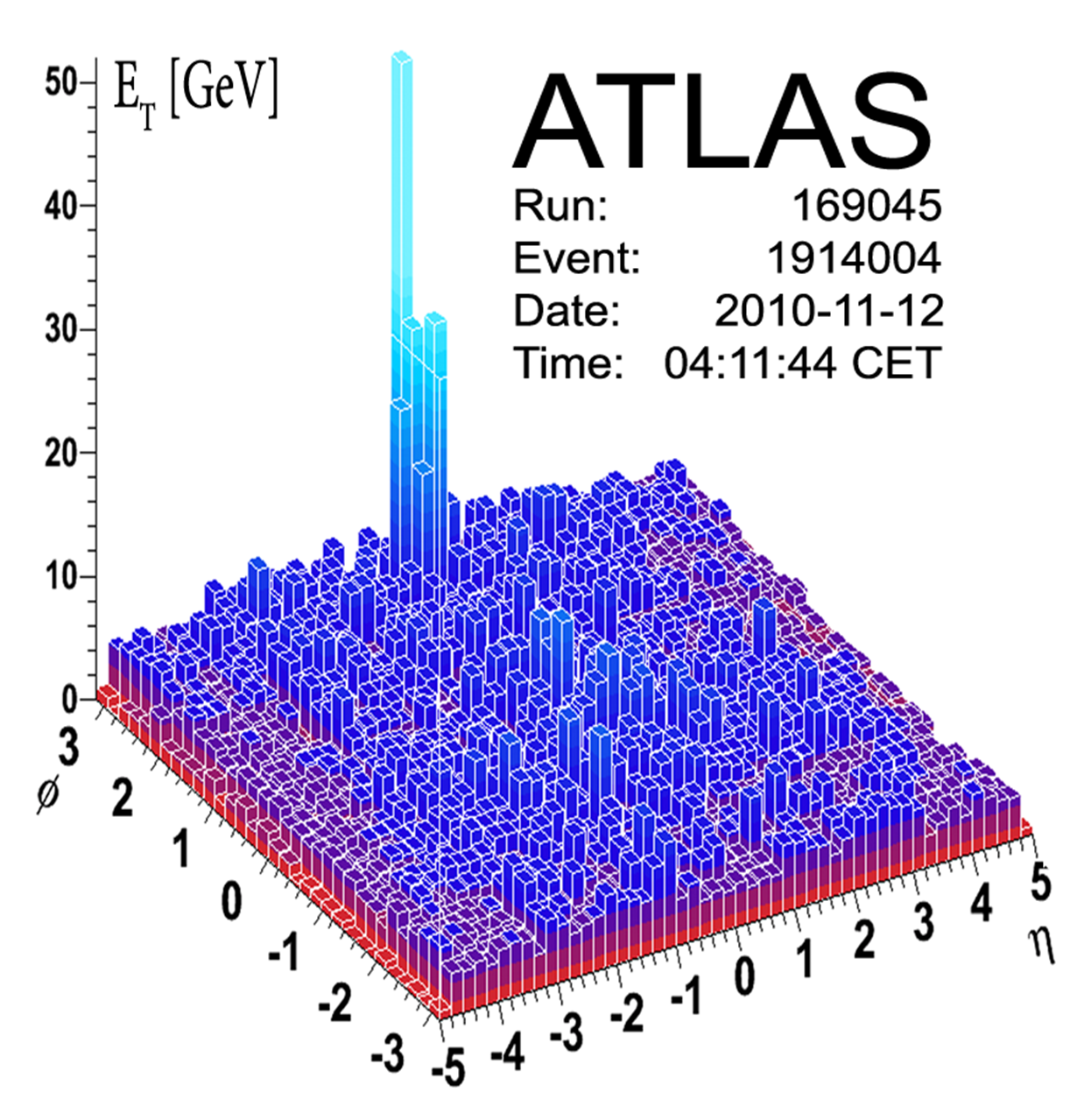}} 
\caption{[Calorimeter display of a very asymmetric (quenched) two-jet event~\cite{Aad:2010bu}.}
\label{LHCAtlasjet}
\end{minipage} & \begin{minipage}{.52\textwidth}
\centerline{\includegraphics[width=1.0\textwidth]{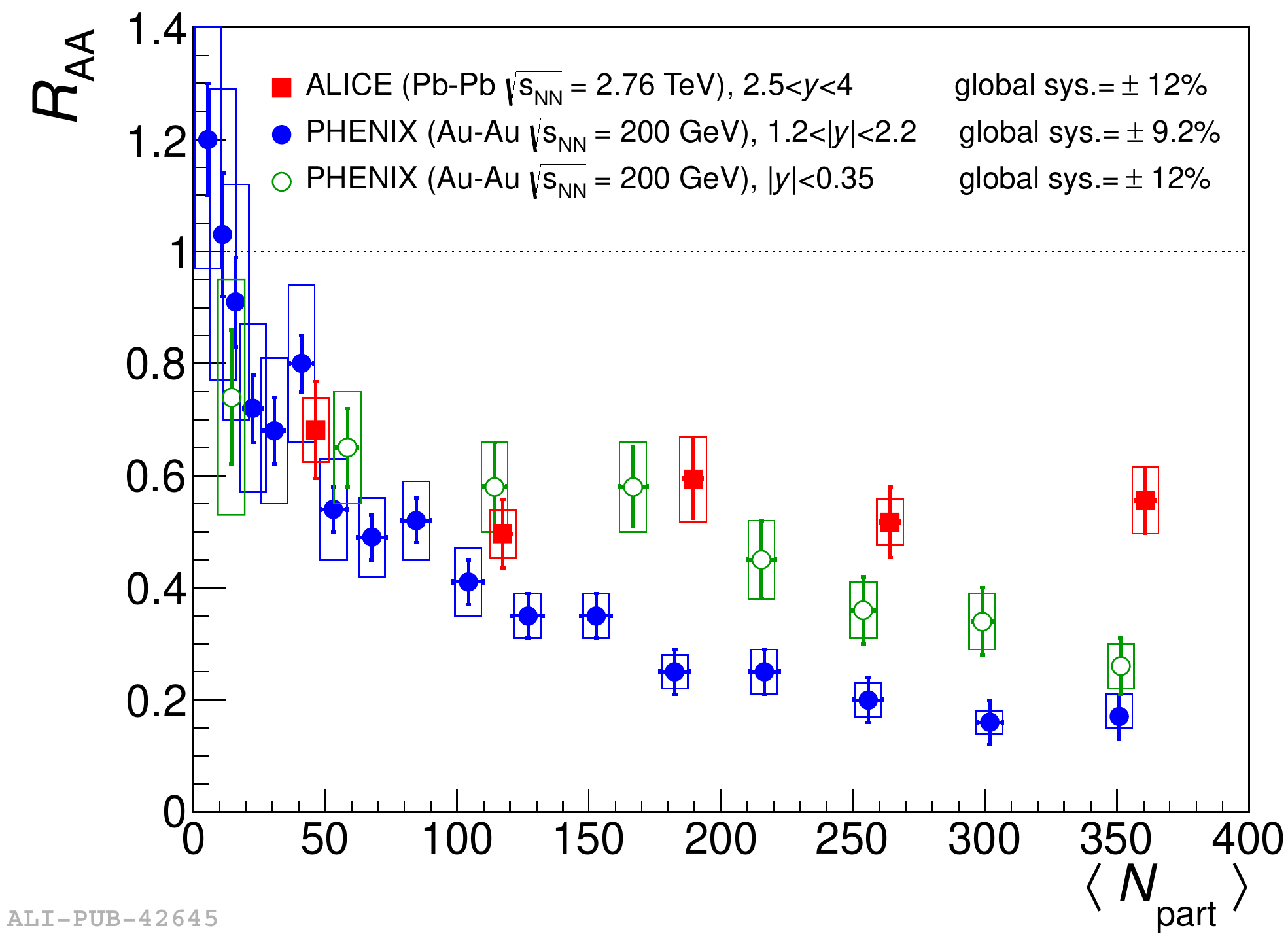}}
\caption{[\Jpsi~suppression versus centrality at RHIC and LHC~\cite{Abelev:2012rv}; $N_{part}$ is the number of participating nucleons which increases with increasing centrality (decreasing impact parameter).}
\label{LHCJpsiRaa}
\end{minipage}
\end{tabular}
\end{figure}

 Jet quenching was discovered at RHIC not with jets, which are difficult to measure in the high multiplicity heavy ion background environment, but as a suppression of high \npt~'leading' jet-fragments. The effect was experimentally very clean and significant with suppression factors up to five.
 The high energy of LHC and the correspondingly large cross sections for hard processes make high energy jets easily stand out from the background even in central nuclear collisions(Fig.~\ref{LHCAtlasjet}). Jet quenching is therefore readily recognized and measured, with many unbalanced dijets or even monojets apparent in the data~\cite{Aad:2010bu}. While the amount of energy lost in the medium can be of the order of tens of GeV and therefore even on average corresponds to a sizeable fraction of the total jet energy, it is nevertheless close to the one expected when extrapolating RHIC results to the higher density matter at LHC. The two jets remain essentially back-to-back (little or no angular broadening relative to pp) and the radiated energy ($\Delta E$) is found in very low \npt~particles ($< 2$ \Gc) and at large angles to the jet direction~\cite{Chatrchyan:2011sx}. The latter two findings were initially a surprise, but are now incorporated naturally into models where the energy is lost in multiple, soft scatterings, and the radiated gluons are emitted at large angles. The parton then leaves the matter and undergoes normal vacuum fragmentation, i.e. looking like a normal $pp$ jet but with a reduced energy.
 
 Additional insight into the energy loss process has come from heavy flavours~\cite{ Renk:2013kva,Dainese:2013vka}. The suppression of charm mesons is virtually identical to the one of inclusive charged particles; a result which was counterintuitive and initially confusing. The similarity in the energy loss of gluons (the source of the majority of charged particles) and heavy quarks is now understood as an accidental cancellation between the difference in coupling strength (colour charge) of quarks and gluons and their different fragmentation functions. The mass effect however seems to be as predicted: At intermediate \npt, beauty shows less suppression than charm, whereas at very high \npt b-jets and inclusive jets show similar modifications.
 
 \subsection{Quarkonium suppression} \label{jpsi}

 While the 'anomalous' $J/\Psi$ suppression discovered at the SPS was considered one of the strongest indications for the QGP, the RHIC results showed essentially the same suppression at a much higher energy, contrary to most expectations and predictions from both QGP and non-QGP models. These initially very confusing results kept the interpretation of this most direct signal for deconfinement ambiguous for the last 10 years.
 
 It had been suggested that \Jpsi~suppression actually increases with energy (i.e. from SPS to RHIC), but is more or less balanced by a new production mechanism: upon reaching the parton-hadron phase boundary two independently produced charm quarks from the plasma hadronize along with the lighter quarks, forming J/Psi~\cite{BraunMunzinger:2009ih}. And indeed, LHC data seems to have resolved the \Jpsi~puzzle in favour of this coalescence picture~\cite{Abelev:2012rv}: as predicted, the large charm cross section at LHC leads to \emph{less} \Jpsi~suppression at LHC compared to RHIC (Fig.~\ref{LHCJpsiRaa}). The suppression is also less strong at low \npt, where phase space favours recombination, in clear contrast to the opposite \npt~dependence found at SPS and RHIC.
 
 While at first sight charm quark coalescence may appear as yet another process complicating and masking quarkonium deconfinement, it is actually a respectable and important deconfinement signal in itself: only in a colour conducting, deconfining medium can quarks roam freely over large distances ($\gg$ 1 fm), and this is exactly what two charm quarks have to do in order to combine during hadronisation.
 
 The magnitude of the suppression for different quarkonium states should depend on their binding energy, with strongly bound states such as the $\Upsilon$ showing less or no modification. LHC results for the $\Upsilon$ family~\cite{Chatrchyan:2011pe} are fully consistent with the expectation for a deconfining hot medium in which quarkonia survival decreases with binding energy, i.e. in terms of suppression factors: \nY(3S) $>$ \nY(2S) $>$ \nY(1S). The \nY(1S) is suppressed by about a factor of two in central collisions, the \nY(2S) by almost an order of magnitude, and only upper limits have been measured for the \nY(3S). As only about 50\% of the observed \nY(1S) are directly produced, these results may be compatible with almost complete melting of all high mass bottonium states and survival of a lone, strongly bound \nY(1S), which according to lattice QCD may melt only at temperatures far above the critical temperature.

 \subsection{Discoveries} \label{ridge}

\begin{figure}[htb]
 \centerline{\includegraphics [width=1.0\textwidth] {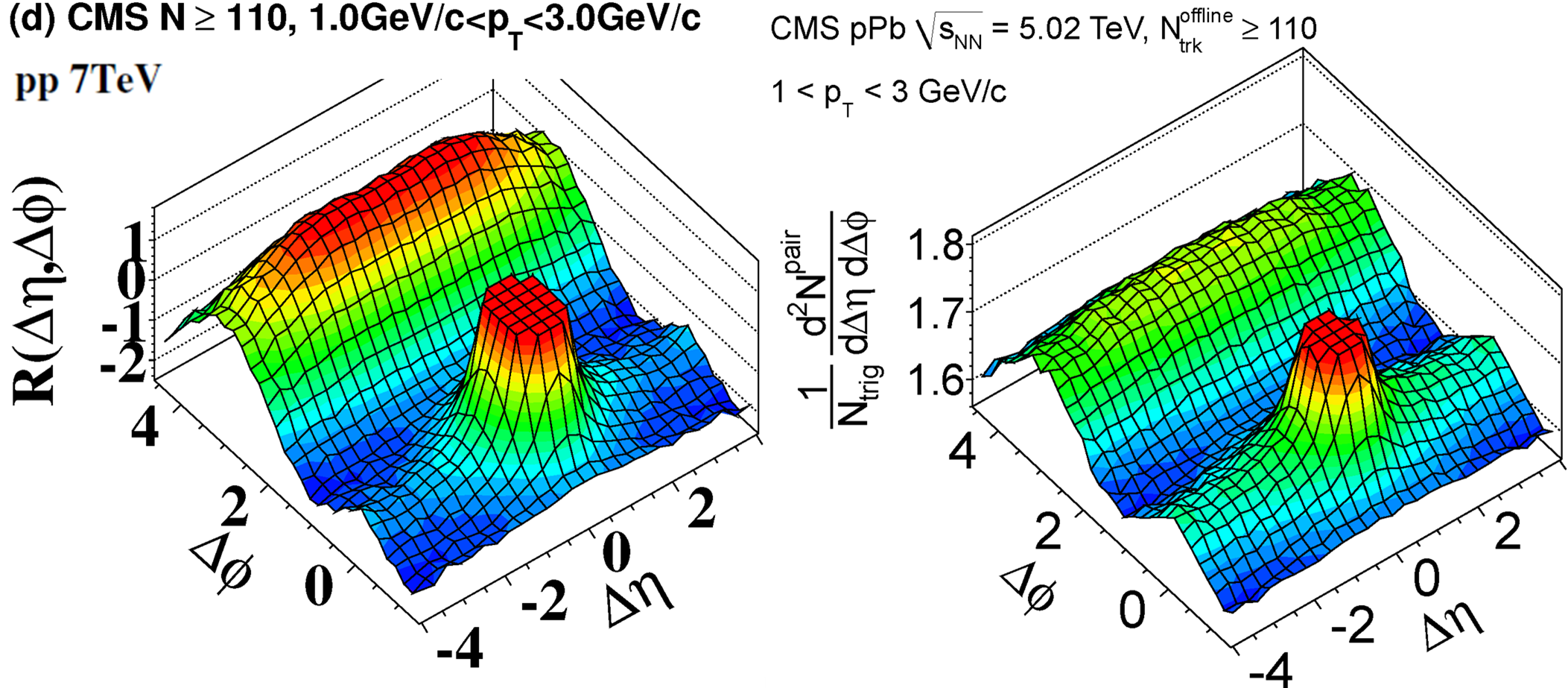}}
 \caption{Two particle correlation function in $\eta-\phi$ for high multiplicity pp collisions~\cite{Khachatryan:2010gv} (left) and pPb collisions~\cite{CMS:2012qk}(right).}
 \label{LHCCMSRidge}
 \end{figure}

 The first discovery made at LHC was announced~\cite{cmsseminar} in Sept. 2010 on a subject which was as unlikely as it was unfamiliar to most in the packed audience: The CMS experiment had found a mysterious 'long range rapidity correlation' in a tiny subset of extremely high multiplicity $pp$ collisions at 7 TeV~\cite{Khachatryan:2010gv}. The correlation in rapidity $\Delta\eta$ and azimuthal angle $\Delta\phi$ between all pairs of particles of intermediate \npt~($1 - 3$ GeV/$c$) in pp collisions is shown in Fig.~\ref{LHCCMSRidge} left. Besides the so called 'near side peak' at (0,0,), a feature arising from particle correlations within jets, and the 'away side ridge' at $\Delta\phi = \pi$ in azimuth, where the two particles come - one each - from the members of a pair of back-to-back jets, the correlation structure shows a small but significant second ridge also at $\Delta\phi = 0$.
 While in the meantime far eclipsed by the discovery of a Higgs particle, this 'near side ridge' is arguably still the most unexpected LHC discovery to date and spawned a large variety of different explanations~\cite{Li:2012hc}. The most serious contenders are saturation physics, as formulated in the Colour Glass Condensate model (CGC)~\cite{Gelis:2010nm}, and collective hydrodynamic flow. Hydrodynamics is of course a very successful framework to describe long range correlations in the macroscopic hot matter created in heavy ion reactions, but was not supposed to be applicable in small systems like $pp$ collisions, where typically only a few ten particles are produced per unit of rapidity. The CGC is a 'first principles' classical field theory approximation to QCD which is applicable to very dense (high occupation number) parton systems like those found at small-x and small $Q^2$ in the initial state wave function of hadrons. It has been successfully used to describe some regularities seen e.g. in ep collisions at HERA ('geometric scaling') and to model the initial conditions in heavy ion physics.

 Lacking further experimental input, no real progress was made to unravel the origin of these long range $pp$ correlations until the ridge made a robust come-back with the first LHC proton-nucleus run some two years later~\cite{CMS:2012qk} (Fig.~\ref{LHCCMSRidge} right, p-Pb at $\sqrt{s_{NN}} = 5$ TeV). The correlation strength was actually significantly stronger than in $pp$ at the same multiplicity, and in quick succession it was discovered that~\cite{ Loizides:2013nka}: the ridge was actually double-sided, showing correlations between particles both close by in azimuth as well as back-to-back; a Fourier analysis revealed both even ($v_2$) as well as odd ($v_3$) components; the dependence of the correlation strength on particle mass was virtually identical to the one expected from hydrodynamic flow; and finally the correlation strength measured with multi-particle methods was almost identical to the one measured with two particles only, convincingly demonstrating that the ridge is a true collective effect which involves \emph{all} low energy particles in \emph{every} event (in contrast to e.g. jet correlations, which involve only a \emph{few} particles in \emph{some} events).
 
 All characteristics of the p-Pb ridge are very natural for and in good agreement with a hydrodynamic collective flow origin of the correlation. Even the strength of the signal and its multiplicity dependence are of the correct order of magnitude (within a factor of two) if one uses some reasonable geometrical initial conditions and a standard hydro model and just \emph{postulates} that the tiny and very short-lived interacting matter system, some 1 fm in size and lifetime, behaves like a macroscopic ideal fluid. Note that the matter created in central Pb-Pb collisions has a size of the order of 5000 fm$^3$ and therefore is larger by orders of magnitude!
 
 The question how such a tiny (few fm$^3$) system could thermalize in essentially no time, maybe even become a small droplet of sQGP, has kept the case open, despite what looks like convincing evidence, including very recent and spectacular confirmation of the ridge effect at RHIC using 'elliptical' deuteron projectiles and 'triangular' $^3$He nuclei.
 
 In any case, the ridge discovery in $pp$ and $pA$ at LHC is definitely more than a curiosity and likely to have profound implications for heavy ion physics. If a sQGP (like) state can be created and studied in much smaller systems than anticipated, we can compare $pp, pA$, and $AA$ to look for finite size effects, which may reveal information on correlation lengths and relaxation time scales not otherwise easily available. If, on the contrary, initial state effects and saturation physics are the answer, we would have discovered at LHC yet another new state of matter, the Colour Glass Condensate, opening a rich new field of activity for both experiment and theory.
 
 \section{Conclusions}
 CERN has been an essential player and (mostly) unwavering supporter in the genesis and advance of high energy heavy ion physics. In the incredibly short time span of little over 30 years, the study of the phases of nuclear matter has evolved from light ion reactions at a fixed target energy of some GeV/nucleon to using heavy projectiles at a center-of-mass energy of several TeV/nucleon, increasing the available energy by three orders of magnitude~\cite{Schukraft:2006nt}.
 This rapid progress was of course only possible by reusing machines, and initially even detectors, built over a longer time scale for particle physics. Today, with more than 2000 physicists active worldwide in this field, ultra-relativistic heavy ion physics has moved in less than a generation from the periphery into a central activity of contemporary Nuclear Physics.
 From the early exploratory phase, with sometimes more qualitative than quantitative results and conclusions, the field has grown up and matured, making important and often unexpected discoveries at each new facility. The view of the Quark-Gluon Plasma has dramatically advanced, from a simple weakly interacting parton gas to a strongly interacting ideal fluid that might find a field theoretical description in a so-called "dual", string theoretical framework~\cite{CasalderreySolana:2011us}.

Today, at its height, heavy ion physics has found interest well beyond the circle of its immediate practitioners, with links and cross fertilisation towards neighbouring disciplines ranging from plasma physics to string theory.
The heavy ion program is very active and competitive today at both high and low energy, to map the phase diagram, locate the transition between normal matter and the sQGP, and to search for a conjectured 'tri-critical' point somewhere in the region at or below SPS fixed target energy. Two new low energy facilities (FAIR at GSI and NICA at JINR) are being built to study compressed matter, i.e. matter at high baryon density and (comparatively) low temperature where the phase structure may be quite different ($1^{st}$ order phase transition) and the matter is closer related to neutron stars than to the early universe.
 The LHC however is and will be the energy frontier facility not only of high energy physics but also of nuclear physics for the foreseeable future, with a well-defined and extensive program and wish list of measurements. And if the first three years can be a guide, strong interaction physics, while firmly rooted in the Standard Model, has shown no end to surprises and discoveries and promises to keep physics with heavy ions interesting (and fun) for quite some time to come.

 \end{document}